\documentclass{article}
\usepackage{arxiv}
\usepackage{cite}
\usepackage{amsmath,amssymb,amsfonts}
\usepackage{algorithmic}
\usepackage{graphicx}
\usepackage{textcomp}
\usepackage{xcolor}
\usepackage{graphicx}
\usepackage{listings}
\usepackage{subcaption}
\usepackage{booktabs}
\usepackage{xcolor}
\usepackage{color,soul}
\usepackage{todonotes}
\def\BibTeX{{\rm B\kern-.05em{\sc i\kern-.025em b}\kern-.08em
    T\kern-.1667em\lower.7ex\hbox{E}\kern-.125emX}}
\usepackage[hidelinks]{hyperref}

\usepackage{todonotes}
\usepackage[numbers,sort&compress,sectionbib]{natbib}
\usepackage{amssymb}
\definecolor{darkgreen}{rgb}{0.1, 0.7, 0.3}
\usepackage{listings}

\usepackage{verbatim}

\usepackage{scalerel}
\usepackage{tikz}
\usetikzlibrary{svg.path}

\definecolor{orcidlogocol}{HTML}{A6CE39}
\tikzset{
  orcidlogo/.pic={
    \fill[orcidlogocol] svg{M256,128c0,70.7-57.3,128-128,128C57.3,256,0,198.7,0,128C0,57.3,57.3,0,128,0C198.7,0,256,57.3,256,128z};
    \fill[white] svg{M86.3,186.2H70.9V79.1h15.4v48.4V186.2z}
                 svg{M108.9,79.1h41.6c39.6,0,57,28.3,57,53.6c0,27.5-21.5,53.6-56.8,53.6h-41.8V79.1z M124.3,172.4h24.5c34.9,0,42.9-26.5,42.9-39.7c0-21.5-13.7-39.7-43.7-39.7h-23.7V172.4z}
                 svg{M88.7,56.8c0,5.5-4.5,10.1-10.1,10.1c-5.6,0-10.1-4.6-10.1-10.1c0-5.6,4.5-10.1,10.1-10.1C84.2,46.7,88.7,51.3,88.7,56.8z};
  }
}

\newcommand\orcidicon[1]{\href{https://orcid.org/#1}{\mbox{\scalerel*{
\begin{tikzpicture}[yscale=-1,transform shape]
\pic{orcidlogo};
\end{tikzpicture}
}{|}}}}

\begin{document}

\title{Supporting OpenMP 5.0 Tasks in hpxMP - A study of an OpenMP implementation within Task Based Runtime Systems}

\author{
  Tianyi Zhang, Shahrzad Shirzad, Bibek Wagle, Adrian S. Lemoine, Patrick Diehl\orcidicon{0000-0003-3922-8419}{}, and Hartmut Kaiser  \\
  Center of Computation \& Technology\\
  Louisiana State University\\
  Baton Rouge, LA \\
  \texttt{\{tzhan18,sshirz1,bwagle3\}@lsu.edu,\{aserio,pdiehl,hkaiser\}@cct.lsu.edu} 
}

\maketitle

\begin{abstract}
OpenMP has been the de facto standard for single node parallelism for more than a decade. Recently, asynchronous many-task runtime (AMT) systems have increased in popularity as a new programming paradigm for high performance computing applications. One of the major challenges of this new paradigm is the incompatibility of the OpenMP thread model and other AMTs. Highly optimized OpenMP-based libraries do not perform well when coupled with AMTs because the threading of both libraries will compete for resources. This paper is a follow-up paper on the fundamental implementation of hpxMP, an implementation of the OpenMP standard which utilizes the C++ standard library for Parallelism and Concurrency (HPX) to schedule and manage tasks~\cite{2019arXiv190303023Z}. In this paper, we present the implementation of task features, e.g. \lstinline|taskgroup|, \lstinline|task depend|, and \lstinline|task_reduction|, of the OpenMP 5.0 standard and optimization of the \lstinline|#pragma omp parallel for| pragma. We use the daxpy benchmark, the Barcelona OpenMP Tasks Suite, Parallel research kernels, and OpenBLAS benchmarks to compare the different OpenMp implementations: hpxMP, llvm-OpenMP, and GOMP.
We conclude that hpxMP is one possibility to overcome the competition for resources of the different thread models by providing a subset of the OpenMP features using HPX threads. However, the overall performance of hpxMP is not yet comparable with legacy libraries, which are highly optimized for a different programming paradigm and optimized over a decode by many contributors and compiler vendors.
\end{abstract}

\keywords{
OpenMP,hpxMP,Asynchronous Many-task Systems,C++,clang,gcc,HPX
}
%
%
\section{Introduction}
Asynchronous many-task (AMT) systems have emerged as a new
programming paradigm in the high performance computing (HPC) community~\cite{thoman2018taxonomy}.
Many of these applications would benefit from the highly optimized OpenMP-based linear algebra libraries currently available. e.g. eigen, blaze, Intel MKL. However, there is a gap between OpenMP and AMT systems, since the user level threads of the AMT systems interfere with the system threads of OpenMP preventing efficient execution of the application.

To close this gap, this paper introduces hpxMP, an implementation of the OpenMP standard, which utilizes a C++ standard library for parallelism and concurrency (HPX)~\cite{heller2017hpx} to schedule and manage tasks. HpxMP implements the OpenMP standard conform \lstinline|#pragma omp parallel for| pragma~\cite{2019arXiv190303023Z}. Furthermore, the OpenMP standard has, since OpenMP $3.0$~\cite{openmp08}, begun to introduce task-based concepts such as depended tasks and task groups (OpenMP 4.0 \cite{openmp13}), task-loops (OpenMP 4.5~\cite{openmp15}), and detached tasks (OpenMP 5.0 \cite{openmp18}). This work extends hpxMP with the \lstinline|#pragma omp task| pragma to provide the concept of task-based programming and validates its implementation against the Barcelona OpenMP Tasks Suite. Next, hpxMP is validated against the daxpy benchmark, OpenBLAS, and Parallel Research Kernels benchmarks to compare the different OpenMP implementations: hpxMP, llvm-OpenMP, and GOMP.

This paper is structured as follows: Section~\ref{sec:relwork} covers the the related work. Section~\ref{sec:hpx} briefly introduces the features of HPX utilized for the implementation of tasks within hpxMP. Section~\ref{sec:tasks} emphasizes the implementation of OpenMP tasks within the HPX framework and shows the subset of OpenMP standard features implemented within hpxMP. Section~\ref{sec:benchmark} shows the comparison of the different OpenMP implementations and finally, Section~\ref{sec:outlook} concludes the work.

\section{Related Work}
\label{sec:relwork}

\subsection*{Multi-threading solutions }
For the exploitation of shared memory parallelism on multi-core processors many solutions are available and have been intensively studied. The most language independent one is the POSIX thread execution model~\cite{pthreads} which exposes fine grain parallelism. In addition, there are more abstract library solutions like Intel's Threading Building Blocks (TBB)~\cite{inteltbb}, Microsoft's Parallel Pattern Library (PPL)~\cite{microsoftppl}, and Kokkos~\cite{CarterEdwards20143202}. TBB provides task parallelism using a C++ template library. PPL provides in addition features like parallel algorithms and containers. Kokkos provides a common C++ interface for parallel programming models, like CUDA and pthreads.
Programming languages such as Chapel~\cite{chamberlain07parallelprogrammability}
provide parallel data and task abstractions. Cilk~\cite{cilk++} extends the
C/C++ language with parallel loop and fork-join constructs to provide single node parallelism. Fore a very detailed review we refer to~\cite{thoman2018taxonomy}.

With the OpenMP 3.0 standard~\cite{openmp08} the concept of task-based programming was added. The OpenMP 3.1~\cite{openmp11} standard introduced task optimization. Depend tasks and task groups provided by the OpenMP 4.0 standard~\cite{openmp13} improved the synchronization of tasks. The OpenMP 4.5~\cite{openmp15} standard allows task-based loops and the OpenMP 5.0~\cite{openmp18} standard allows detaching of tasks, respectively.

\subsection*{Integration of Multi-threading solutions within distributed programming models.}
Some major research in the area of \textit{MPI+X}~\cite{10.1109/MC.2016.232,barrett2015toward,rabenseifner2009hybrid,smith2001development}, where the Message Passing Interface (MPI) is used as the distributed programming model and OpenMP as the multi-threaded shared memory programming model has be done. However, less research has be done for \textit{AMT+X}, where asynchronous many-task systems are used as the distributed programming model and OpenMP as the shared memory programming model. Charm++ integrated OpenMP's shared memory parallelism to its distributed programming model for improving the load balancing~\cite{bak2017integrating}.  Kstar, a research C/C++ OpenMP compiler~\cite{agullo2017bridging}, was utilized to generate code compatible with StarPU~\cite{augonnet2011starpu} and Kaapi~\cite{gautier2007kaapi}. Only Kaapi implements the full set of OpenMP specification~\cite{broquedis2012libkomp}, such as the capability to create task into the context of a task region. The successor XKaapi~\cite{gautier2013xkaapi} provides C++ task based interface for both multi-core and multi-GPUs and only Kaapi provides mult-cluster support.

\section{C++ Standard Library for Concurrency and Parallelism (HPX)}
\label{sec:hpx}
HPX is an open source C++ standard conferment library for parallelism and concurrency for applications of any scale. One specialty of HPX is that its API offers a syntactic and semantic equivalent interface for local and remote function calls. HPX incorporates  well known concepts, \emph{e.g.} static and dynamic dataflows, fine-grained
futures-based synchronization, and continuation-style programming~\cite{hpx_pgas_2014}. Fore more details we refer to~\cite{heller:2012,Heller:2013:UHL:2530268.2530269,hpx_pgas_2014,Kaiser:2015:HPL:2832241.2832244,hartmut_kaiser_2018_1484427,heller:hpc_2016,wagle2018methodology}.

After this brief overview, let us look into the relevant parts of HPX in the context of this work. HPX is an implementation of the ParalleX execution model~\cite{scaling_impaired_apps} and generates hundreds of millions of so-called light-weight HPX-threads (tasks). These light-weight threads provide fast context switching~\cite{heller2017hpx} and lower overloads per thread to make it feasible to schedule a large number of tasks with negligible overhead~\cite{wagle2019runtime}. However, these light-weighted HPX threads are not compatible with the user threads utilized by the current OpenMP implementation. For more details we refer to our previous paper~\cite{2019arXiv190303023Z}. This works adds support of OpenMP 5.0 task features to hpxMP. With having a light-weighted HPX thread-based implementation of the OpenMP standard it enables applications that already use HPX for local and distributed computations to integrate highly optimized libraries that rely on HPX in future.

\section{Implementation of hpxMP}
\label{sec:tasks}
hpxMP is an implementation of the OpenMP standard, which utilizes a C++ standard library for parallelism and concurrency (HPX)~\cite{heller2017hpx} to schedule and manage threads and tasks. We have described the fundamental implementation of hpxMP in previous work~\cite{2019arXiv190303023Z}. This section addresses the implementation of few important classes in hpxMP, task features, such as \lstinline|taskgroup|, \lstinline|task depend|, and \lstinline|task_reduction| in the OpenMP $5$.$0$ standard~\cite{openmp18}, within hpxMP and its optimization for the thread and task synchronization which are the new contribution of this work.
\subsection{Class Implementation}
An instance of \lstinline|omp_task_data| class is set to be associated with each HPX thread by calling \lstinline|hpx::threads::set_thread_data|. Instances of \lstinline|omp_task_data| are passed by a raw pointer which is \lstinline|reinterpret_cast|ed to \lstinline|size_t|. For better memory management, a smart pointer \lstinline|boost::intrusive_ptr| is introduced to wrap around \lstinline|omp_task_data|. The class \lstinline|omp_task_data| consists the information describing a thread, such as a pointer to the current \lstinline|team|, \lstinline|taskLatch| for synchronization and if the \lstinline|task| is in \lstinline|taskgroup|. The \lstinline|omp_task_data| can be retrieved by calling \lstinline|hpx::threads::get_thread_data| when needed, which plays an important role in hpxMP runtime.

Another important class is \lstinline|parallel_region|, containing information in a \lstinline|team|, such as \lstinline|teamTaskLatch| for task synchronization, number of threads requested under the parallel region, and the depth of the current team.
\subsection{Task Construct}
Explicit tasks are created using the task construct in hpxMP. hpxMP has
implemented the most recent OpenMP 5.0 tasking features and synchronization
constructs, like \lstinline|task|, \lstinline|taskyield|, and \lstinline|taskwait|.
The supported clause associated with \lstinline|#pragma omp task|
are \lstinline|reduction|, \lstinline|untied|, \lstinline|private|,
\lstinline|firstprivate|, \lstinline|shared|, and \lstinline|depend|.

Explicit tasks are created using \lstinline|#pragma omp task| in hpxMP. HPX threads are created with the task directives and tasks are running on these HPX threads created. \lstinline|_kmpc_omp_task_alloc| is allocating, initializing tasks and then return the generated tasks to the runtime.
\lstinline|__kmpc_omp_task| is called with the generated \lstinline|task| parameter and passed to the \lstinline|hpx_runtime::create_task|. The tasks are then running as a normal priority HPX thread by calling function \lstinline|hpx::applier::register_thread_nullary|, see Listing~\ref{lst:task}. Synchronization in tasking implementation of hpxMP are handled with HPX latch,  which will be discussed later in Section~\ref{sec:thr_sync}.

Task dependency was introduced with OpenMP 4.0. The depend clause is \lstinline|#pragma omp task depend(in: x) depend( out: y) depend(inout: z)|. Certain dependency should be satisfied among tasks specified by users. In the implementation, \lstinline|future| in HPX is employed. The functionality called \lstinline|hpx::future| allows for the separation of the initiation of an operation and the waiting for the result. A list of tasks that current task depend on are stored in a \lstinline|vector<shared_future<void>>| and \lstinline|hpx::when_all(dep_futures)| are called to inform the current task when it is ready to run.

OpenMP 5.0 added great extension to the tasking structure in OpenMP. \lstinline|task_reduction| along with \lstinline|in_reduction| gives users a way to tell the compiler reduction relations among tasks and specify the tasks in taskgroup which are participating the reduction. The implementation of taskgroup can be found in Listing~\ref{lst:taskgroup}. Reduction data is handled in
\lstinline|kmpc_task_reduction_init|, by assigning them to the taskgroups, and return the taskgroup data back to the runtime.
\lstinline|#pragma omp task in_reduction ( operator : list )|
tells the runtime which task is participating the reduction, and retrieves the reduction data by calling \lstinline|__kmpc_task_reduction_get_th_data|.
\lstinline|kmp_task_reduction_fini| is called by
\lstinline|kmpc_end_taskgroup|, cleaning memory allocated
and finish the task reduction properly.

\begin{lstlisting}[caption=Implementation of task scheduling in hpxMP,label={lst:task},float=*ptb,numbers=left,basicstyle=\scriptsize\ttfamily]
kmp_task_t* __kmpc_omp_task_alloc(... )
{
	kmp_task_t *task = (kmp_task_t*)new char[task_size + sizeof_shareds];
//lots of initilization goes on here
	return task;
}
void hpx_runtime::create_task( kmp_routine_entry_t task_func, int gtid, intrusive_ptr<kmp_task_t> kmp_task_ptr){
	auto current_task_ptr = get_task_data();
//this is waited in taskwait, wait for all tasks before taskwait created to be done
// create_task function is not supposed to wait anything
	current_task_ptr->taskLatch.count_up(1);
//count up number of tasks in this team
	current_task_ptr->team->teamTaskLatch.count_up(1);
//count up number of task in taskgroup if we are under taskgroup construct
	if(current_task_ptr->in_taskgroup)
		current_task_ptr->taskgroupLatch->count_up(1);
//Create a normal priority HPX thread with the allocated task as argument.
	hpx::applier::register_thread_nullary(.....)
	return 1;
}
\end{lstlisting}
\begin{lstlisting}[caption=Implementation of \_\_kmpc\_taskgroup and \_\_kmpc\_end\_taskgroup in hpxMP,label={lst:taskgroup},float=*ptb,numbers=left,,basicstyle=\scriptsize\ttfamily]
void __kmpc_taskgroup( ident_t* loc, int gtid )
{
	auto task = get_task_data();
	intrusive_ptr<kmp_taskgroup_t> tg_new(new kmp_taskgroup_t());
	tg_new->reduce_num_data = 0;
	task->td_taskgroup = tg_new;
	task->in_taskgroup = true;
	task->taskgroupLatch.reset(new latch(1));
}
void __kmpc_end_taskgroup( ident_t* loc, int gtid )
{
	auto task = get_task_data();
	task->tg_exec.reset();
	task->taskgroupLatch->count_down_and_wait();
	task->in_taskgroup = false;
	auto taskgroup = task->td_taskgroup;
	if (taskgroup->reduce_data != NULL)
		__kmp_task_reduction_fini(nullptr,taskgroup);

}
\end{lstlisting}

\subsection{Thread and Task Synchronization}
\label{sec:thr_sync}
In this work we improved the performance of hpxMP over previous
versions~\cite{2019arXiv190303023Z} by optimizing the control structures used for thread synchronization. Previously, an exponential back-off is used for thread synchronization. Now, HPX latch, see Listing~\ref{lst:latch}, provides an easier to use and more efficient way to manage thread and task synchronization originally proposed in the draft C++ library Concurrency Technical specification{\footnote{\url{https://github.com/cplusplus/concurrency-ts/blob/master/latch_barrier.html}}}. Latch in HPX is implemented with mutex, condition variable, and locks however is well-designed and higher level. An internal counter is initialized in a latch to keep track of a calling thread needs to be blocked. The latch blocks one or more threads from executing until the counter reaches 0. Several member functions such as \lstinline|wait()|, \lstinline|count_up()|, \lstinline|count_down()|, \lstinline|count_down_and_wait()| of the Latch class is provided. The difference between \lstinline|count_down()| and \lstinline|count_down_and_wait()| is if the thread will be blocked if the data member inside Latch is not equal to 0 after decreasing the counter by 1.
\begin{lstlisting}[caption=Defination of Latch Class in HPX,label={lst:latch},float=*tb,numbers=left,basicstyle=\scriptsize\ttfamily]
class Latch
{
public:
	void count_down_and_wait();
	void count_down(std::ptrdiff_t n);
	bool is_ready() const noexcept;
	void wait() const;
	void count_up(std::ptrdiff_t n);
	void reset(std::ptrdiff_t n);
protected:
	mutable util::cache_line_data<mutex_type> mtx_;
	mutable util::cache_line_data<local::detail::condition_variable> cond_;
	std::atomic<std::ptrdiff_t> counter_;
	bool notified_;
}
\end{lstlisting}
In parallel regions, when one thread is spawning a team of threads, an HPX latch called threadLatch will be initialized to \lstinline|threads_requested+1| and member function \lstinline|threadLatch.count_down_and_wait()| is called by the parent thread after threads are spawned, making parent threads wait for child threads to finish their work. The Latch is passed as a reference to each child thread and the member function \lstinline|threadLatch.count_down()| is called by each child thread when their works are done. When all the child threads have called the member function, the internal counter of \lstinline|threadLatch| will be reduced to 0 and the thread will be released.
For task synchronization, the implementation is trickier and needs to be carefully designed. In Listing~\ref{lst:task}, three Latches \lstinline|taskLatch|, \lstinline|teamTaskLatch|, and \lstinline|taskgroupLatch| are \lstinline|count_up(1)| when a task is created. Based on the definition of OpenMP standard, tasks are not necessarily synchronized unless a \lstinline|#pragma omp taskwait| or \lstinline|#pragma omp barrier| is called either explicitly or implicitly, see Listing~\ref{lst:taskwait}. The member function of Latch \lstinline|count_down(1)| is called when a task is done with its work. \lstinline|TaskLatch| only matters when \lstinline|#pragma omp taskwait| is specified, where \lstinline|taskLatch.wait()| is called, making sure the current task is suspended until all child tasks that it generated before the taskwait region complete execution. The \lstinline|teamTaskLatch| is used to synchronize all the tasks under a team, including all child tasks this thread created and all of their descendant tasks. An implicit barrier is always triggered at the end of parallel regions, where \lstinline|team->teamTaskLatch.wait()| is called and the current task can be suspended.
\begin{lstlisting}[caption=Implementation of taskwait and barrier wait ,label={lst:taskwait},float=*tb,numbers=left,basicstyle=\scriptsize\ttfamily]
void hpx_runtime::task_wait()
{
	auto task = get_task_data();
	intrusive_ptr<omp_task_data> task_ptr(task);
	task_ptr->taskLatch.wait();
}
void hpx_runtime::barrier_wait()
{
	auto *team = get_team();
	task_wait();
//wait for all child tasks to be done
	team->teamTaskLatch.wait();
}
\end{lstlisting}
Taskgroup implementation in hpxMP is similar to a barrier, see Listing~\ref{lst:taskgroup}. All tasks under the same taskgroup are blocked until the \lstinline|taskgroupLatch->count_down_and_wait()| function inside \lstinline|kmpc_end_taskgroup| is called by all child tasks and their descend tasks.

\subsection{Recap of the implementation}
This sections summarizes the previous presented features of the OpenMP standard implemented with hpxMP. Table~\ref{tab:directive} shows the directives provided by the program layer and correspond to the main part of the presented library. Table~\ref{tab:libfuncs} shows the runtime library functions of the OpenMP standard provided by hpxMP. Of course, the pragmas and runtime library functions are only a subset of the OpenMP specification, but one step to bridge the compatibiltiy gap between OpenMP and the HPX runtime system.
\begin{table}[tbp]
\centering
\begin{tabular}{ll}
\lstinline|#pragma omp atomic| & \lstinline|#pragma omp barrier| \\
\lstinline|#pragma omp critical| & \lstinline|#pragma omp for| \\
\lstinline|#pragma omp master| & \lstinline|#pragma omp ordered| \\
\lstinline|#pragma omp parallel| & \lstinline|#pragma omp section| \\
\lstinline|#pragma omp single| & \lstinline|#pragma omp task depend|
\end{tabular}
\caption{Directives implemented in the program layer of hpxMP. These functions correspond to the main part of the presented library.}
\label{tab:directive}
\end{table}

\begin{table}[tbp]
\centering
\begin{tabular}{ll}
\lstinline|omp_get_dynamic| & \lstinline|omp_get_max_threads| \\
\lstinline|omp_get_num_procs| & \lstinline|omp_get_num_threads| \\
\lstinline|omp_get_thread_num| & \lstinline|omp_get_wtick| \\
\lstinline|omp_get_wtime| & \lstinline|omp_in_parallel| \\
\lstinline|omp_init_lock| & \lstinline|omp_init_nest_lock| \\
\lstinline|omp_set_dynamic| & \lstinline|omp_set_lock| \\
\lstinline|omp_set_nest_lock| & \lstinline|omp_set_num_threads| \\
\lstinline|omp_test_lock| & \lstinline|omp_test_nest_lock| \\
\lstinline|omp_unset_lock| & \lstinline|omp_unset_nest_lock| \\
\end{tabular}
\caption{Directives implemented in the program layer of hpxMP. These functions correspond to the main part of the presented library.}
\label{tab:libfuncs}
\end{table}

\section{Comparison of the OpenMP implementations}
\label{sec:benchmark}
In this paper, the Daxpy Benchmark, Parallel Research Kernels and Barcelona OpenMP Tasks Suite
are used to compare the performance between three different
implementations: hpxMP, llvm-OpenMP, and GOMP, which are provided by
the authors, Intel, and GNU project respectively. The threads are pinned under each measurement for llvm-OpenMP and GOMP. The Blazemark
benchmarks{\footnote{\url{https://bitbucket.org/blaze-lib/blaze/wiki/Benchmarks}}}
from the authors previous work~\cite{2019arXiv190303023Z} are rerun
to emphasize the recent improvements of performance. The benchmarks
are tested on Marvin ($2$ x Intel\textsuperscript{\textregistered}
Xeon\textsuperscript{\textregistered} CPU E5-2450 0 @ 2.10GHz and
$48$ GB RAM), a node having $16$ physical cores in two NUMA domains.

The versions of Clang, GCC, LLVM OpenMP and GOMP used
were 8.0.0, 9.1.0, 4.5 and 4.5 respectively. We used hpxMP with commit id d9234c2, HPX with commit id 414380e, Blaze 3.4\footnote{\url{https://bitbucket.org/blaze-lib/blaze}},
Boost 1.70 and gperftools 2.7. The operating system used was CentOS 7.6.1810
with kernel 3.10.


\subsection{Daxpy Benchmark}
In order to compare the performance of \lstinline|#pragma omp parallel for|, which is a fundamental pragma in OpenMP, Daxpy benchmark is used in this measurement. Daxpy is a benchmark that measures the multiplication of a
float number $c$ with a dense vector $a$ consists 32 bit floating numbers, then add the
result with another dense vector $b$ (32 bit float), the result is stored in the same vector $b$,
where $c\in \mathbb{R}$ and $a,b\in \mathbb{R}^n$.

The Daxpy benchmark compares the performance calculated in
Mega Floating Point Operations Per Second (MFLOP/s).
We determine the speedup of the application by scaling
our results to the single-threaded run of the benchmark using
hpxMP.

Figure~\ref{fig:scale,daxpy} shows the speedup ratio with different
numbers of threads. Our first experiment compared the performance of the OpenMP
implementations when the vector size was set to $10^3$,
see Figure~\ref{fig:ratio,daxpy_d}. llvm-OpenMP runs the fastest while following with GOMP and hpxMP.
Figure~\ref{fig:ratio,daxpy_c} shows that with a
vector size of $10^4$, GOMP and llvm-OpenMP are still able to exhibit
some scaling while hpxMP struggles to scale past $4$ threads.
For very large vector sizes of $10^5$ and $10^6$, the three
implementations perform almost identically. hpxMP is able to
scale in these scenarios because there is sufficient
work in each task in order to amortize the cost of the task
management overheads.

\begin{figure}[p]
	\centering
	\begin{subfigure}[b]{0.35\textwidth}
		\includegraphics[page=1,width=\linewidth]{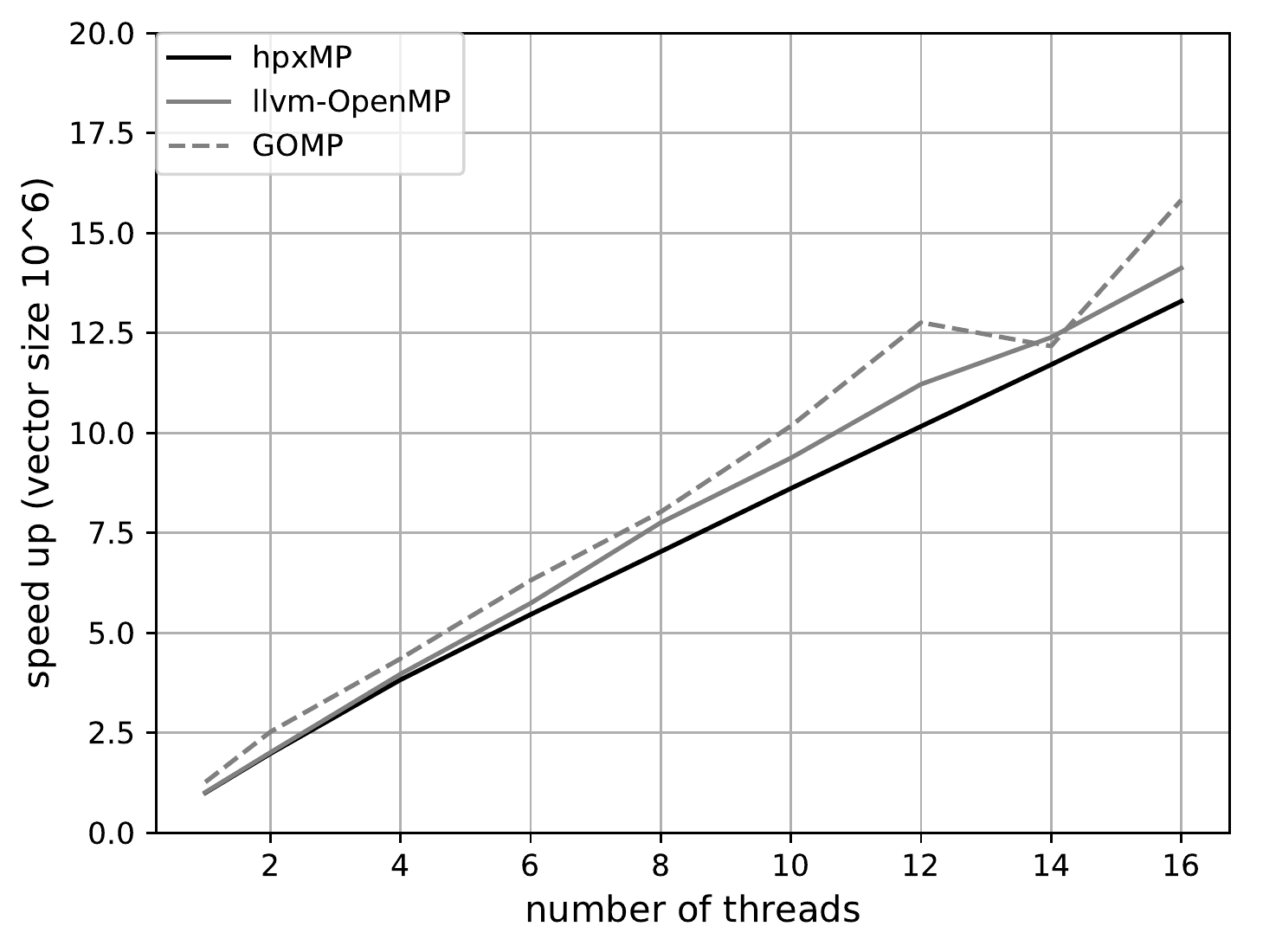}
		\caption{}
		\label{fig:ratio,daxpy_a}
	\end{subfigure}
	\begin{subfigure}[b]{0.35\textwidth}
		\includegraphics[page=3,width=\linewidth]{figure/daxpy/dif_vector_size_speedup_cmb.pdf}
		\caption{}
		\label{fig:ratio,daxpy_b}
	\end{subfigure}
	\begin{subfigure}[b]{0.35\textwidth}
		\includegraphics[page=5,width=\linewidth]{figure/daxpy/dif_vector_size_speedup_cmb.pdf}
		\caption{}
		\label{fig:ratio,daxpy_c}
	\end{subfigure}
	\begin{subfigure}[b]{0.35\textwidth}
		\includegraphics[page=4,width=\linewidth]{figure/daxpy/dif_vector_size_speedup_cmb.pdf}
		\caption{}
		\label{fig:ratio,daxpy_d}
	\end{subfigure}
	\caption{Scaling plots for the Daxpy benchmarks running with different vector sizes: (\subref{fig:ratio,daxpy_a}) $10^6$, (\subref{fig:ratio,daxpy_b}) $10^5$, (\subref{fig:ratio,daxpy_c}) $10^4$, and (\subref{fig:ratio,daxpy_d}) $10^3$. Larger vector sizes mean larger tasks are created. The speedup is calculated by scaling the execution time of a run by the execution time of the single threaded run of hpxMP. A larger speedup factor means a smaller execution time of the sample.}
	\label{fig:scale,daxpy}
\end{figure}
\subsection{DGEMM benchmark}
We chose to use the DGEMM benchmark from Parallel Research Kernels{\footnote{\url{https://github.com/ParRes/Kernels}}}~\cite{7040972} to test our implementation. The purpose of the DGEMM program is to test the performance doing a dense matrix multiplication.
The DGEMM benchmark compares the performance calculated in execution time(seconds).
Figure~\ref{fig:scale,dgemm1} shows the execution time with different
numbers of threads. The performance of the OpenMP implementations when the matrix size was set to $10^3$ is shown in Figure~\ref{fig:scale,dgemm_1}. hpxMP and llvm-OpenMP runs perform similar while both outperform GOMP.
Figure~\ref{fig:scale,dgemm_2} shows that with a matrix size of $100$, GOMP and llvm-OpenMP are still able to exhibit
some scaling while hpxMP struggles to scale past $4$ threads and is slower that GOMP after 8 threads.
\begin{figure}[tbp]
	\centering
	\begin{subfigure}[b]{0.35\textwidth}
		\includegraphics[page=1 ,width=\linewidth]{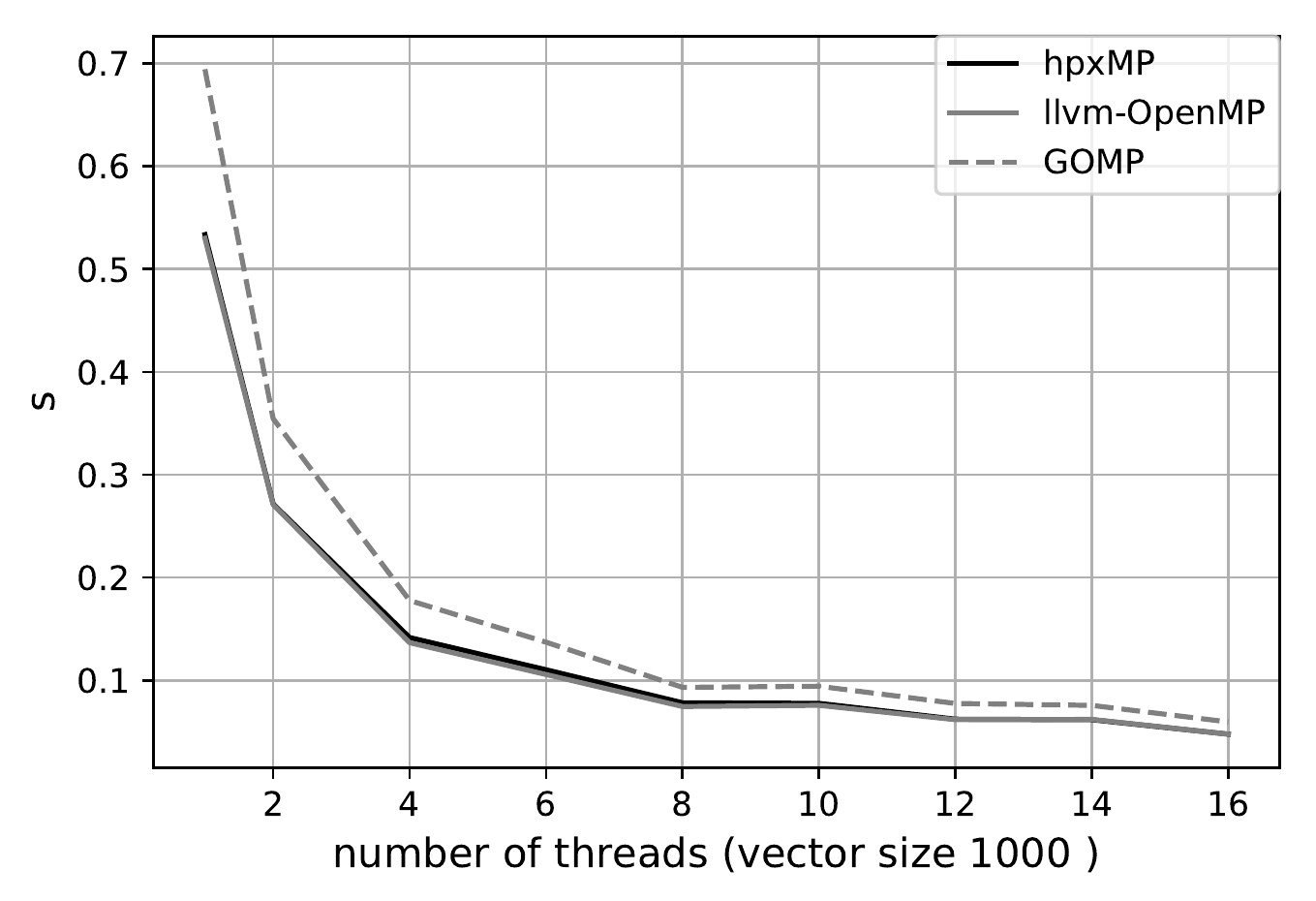}
		\caption{cut off $1000$}
		\label{fig:scale,dgemm_1}
	\end{subfigure}
	\begin{subfigure}[b]{0.35\textwidth}
		\includegraphics[page=3 ,width=\linewidth]{figure/dgemm/dif_vector_size_cmb.pdf}
		\caption{cut off  $100$}
		\label{fig:scale,dgemm_2}
	\end{subfigure}
	\caption{Scaling plots for the Parallel Research Kernels DGEMM Benchmarks ran with the vector size of: (a) $1000$, (b) $100$.  The relation between time consumed and the number of threads using hpxMP, llvm-OpenMP, and GOMP are plotted. The time consumed is calculated by the execution time (s). A smaller time execution time means better performance of the sample.}
	\label{fig:scale,dgemm1}
\end{figure}
\subsection{Barcelona OpenMP Task Suit}
We chose to use the fast parallel sorting variation of the ordinary
mergesort~\cite{akl1987optimal} of the Barcelona OpenMP Tasks Suite
to test our implementation of tasks. We sorted a random array
with $10^7$ 32-bit numbers is sorted with cut off values from $10$ to $10^7$.
The cut off value determines when to perform serial
quicksort instead of dividing the array into $4$ portions recursively
when tasks are created. Higher cut off values create larger size of tasks
and, therefore, fewer tasks are created. In order to
simplify the experiment, parallel merge is disabled and the threshold for
insertion sort is set to $1$ in this benchmark.
For each cut off value, the execution time of hpxMP using $1$
thread is selected as the base point to calculate speedup values.
Figure~\ref{fig:scale,sort1} shows the speedup ratio when using different
numbers of threads.

For the cut off value $10^7$ (Figure~\ref{fig:scale,sort_1}),
the array is divided into four sections and four tasks in total are created.
The speed up curve rapidly increases when moving from $2$ threads
to $4$, but no significant speedup is achieved when using more
than $4$ threads in all three implementations. HpxMP and llvm-OpenMP show
comparable performance while GOMP is
slower.

The cut off value of $10^5$ (Figure~\ref{fig:scale,sort_2}) increases the number of tasks
generated. In this case, llvm-OpenMP has a performance advantage while hpxMP and GOMP show comparable
performance.

For the cut off value $10^3$ (Figure~\ref{fig:scale,sort_3}), llvm-OpenMP shows
a distinct performance advantage over hpxMP and GOMP. Nevertheless,
hpxMP still scales across all the threads while GOMP has ceases to scale past $8$ threads.

For a cut off value of $10$ (Figure~\ref{fig:scale,sort_4}),
a significant number of tasks are created and the work for each task
is considerably small. Here, hpxMP does not scale due to the
the large amount of overheads
associated with the creation of many user tasks.
Because each task performs little work, the overhead that they create
is not amortized by the increase in concurrency.

\begin{figure}[tbp]
	\centering
	\begin{subfigure}[b]{0.35\textwidth}
		\includegraphics[page=3 ,width=\linewidth]{figure/sort/dif_cut_off_speedup_cmb.pdf}
		\caption{cut off $10^7$}
		\label{fig:scale,sort_1}
	\end{subfigure}
	\begin{subfigure}[b]{0.35\textwidth}
		\includegraphics[page=1 ,width=\linewidth]{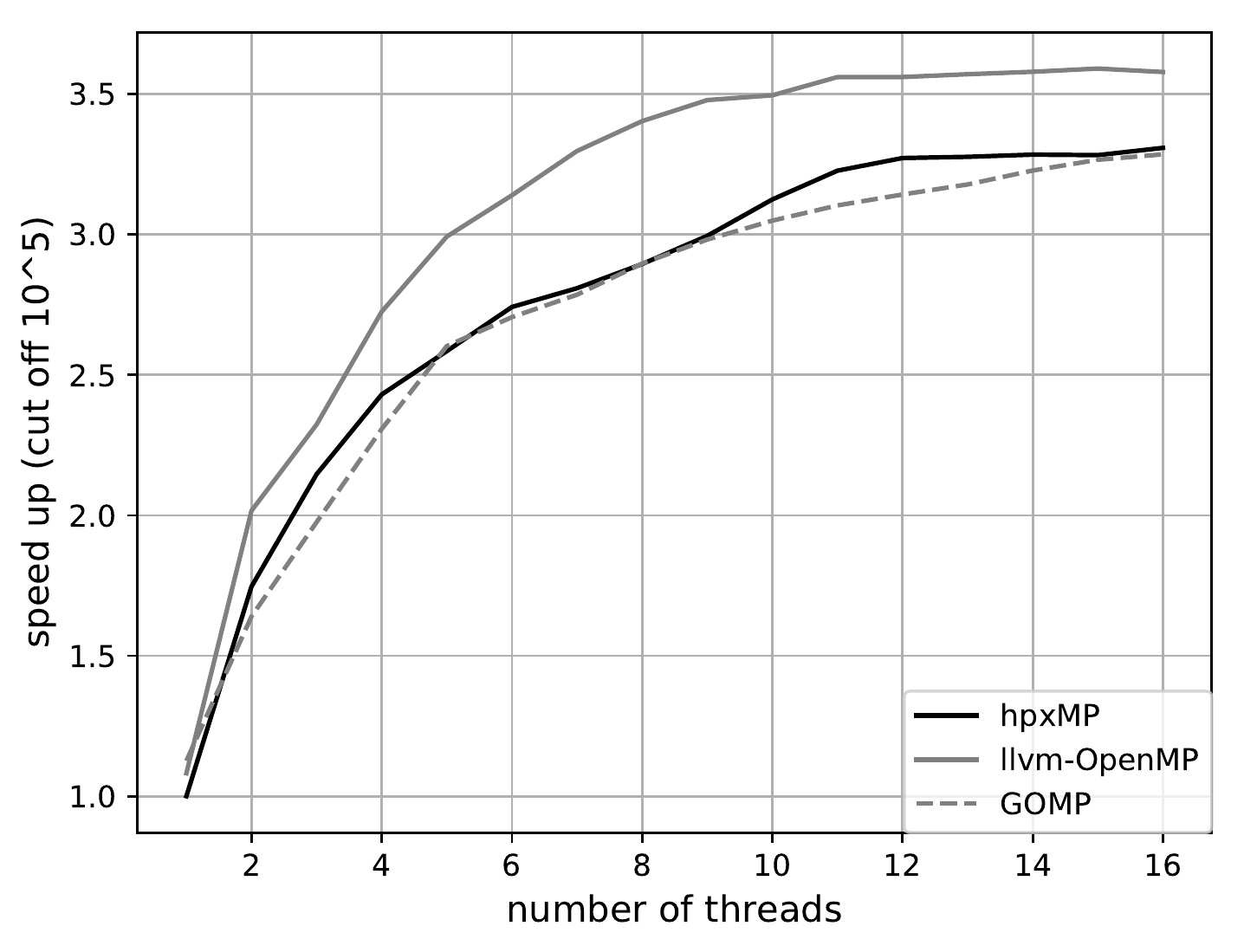}
		\caption{cut off  $10^5$}
		\label{fig:scale,sort_2}
	\end{subfigure}
	\begin{subfigure}[b]{0.35\textwidth}
		\includegraphics[page=5 ,width=\linewidth]{figure/sort/dif_cut_off_speedup_cmb.pdf}
		\caption{cut off  $10^3$}
		\label{fig:scale,sort_3}
	\end{subfigure}
	\begin{subfigure}[b]{0.35\textwidth}
		\includegraphics[page=6 ,width=\linewidth]{figure/sort/dif_cut_off_speedup_cmb.pdf}
		\caption{cut off  $10$}
		\label{fig:scale,sort_4}
	\end{subfigure}
	\caption{Scaling plots for the Barcelona OpenMP Task suit's Sort Benchmarks ran with the cut off values of: (a) $10^7$, (b) $10^5$, (c) $10^3$, and (d) $10$.  Higher cut off values indicate a smaller number of larger tasks are created. The speed up is calculated by scaling the execution time of a run by the execution time of the single threaded run of hpxMP. }
	\label{fig:scale,sort1}
\end{figure}

For a global view, the speedup ratio $r$ is shown in
Figure~\ref{fig:ratio,sort}, where the larger the heatmap value is,
the better performance OpenMP has achieved in comparison to hpxMP.
Values below $1$ mean that hpxMP outperforms the OpenMP implementation.
As shown in the heatmap, llvm-OpenMP works best when the task granularity
is small and the number of tasks created is high. GOMP is slower than both
implementations in most cases. For large task sizes, hpxMP is comparable
with llvm-OpenMP (Figure~\ref{fig:ratio,sort,clang}). This result demonstrates
that when the grain size of the task is chosen well hpxMP will not incur
a performance penalty. Here, some more research has to be done on how hpxMP can handle task granularity and limit the overhead in task management for small grain sizes. Some related work can be found here~\cite{gautier2018impact,vandierendonck2011unified,gautier2013xkaapi,broquedis2012libkomp,bueno2011productive}.

\begin{figure}[tbp]
	\centering
	\begin{subfigure}[b]{0.49\textwidth}
		\includegraphics[width=\linewidth]{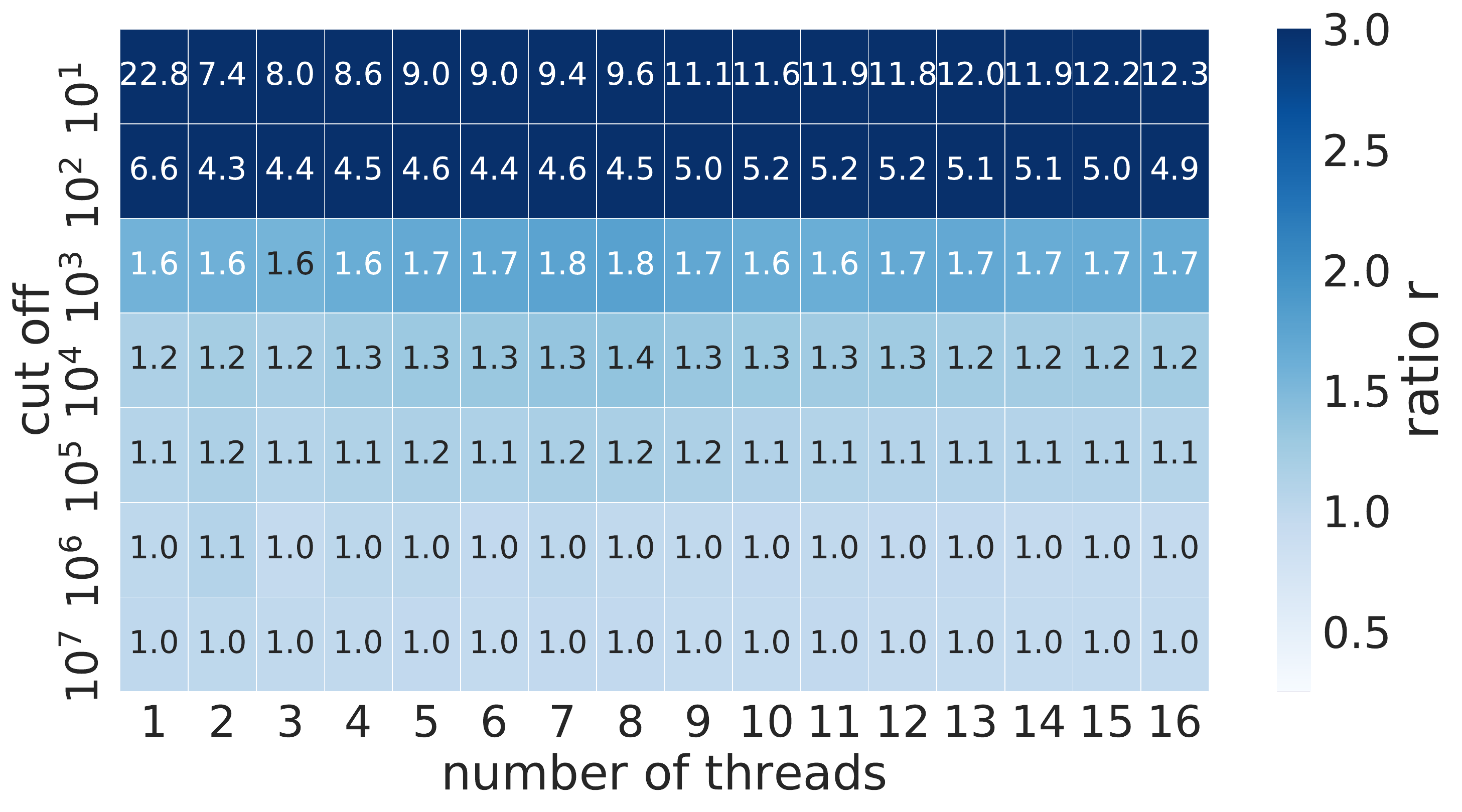}
		\caption{llvm-OpenMP/hpxMP}
		\label{fig:ratio,sort,clang}
	\end{subfigure}
	\begin{subfigure}[b]{0.49\textwidth}
		\includegraphics[width=\linewidth]{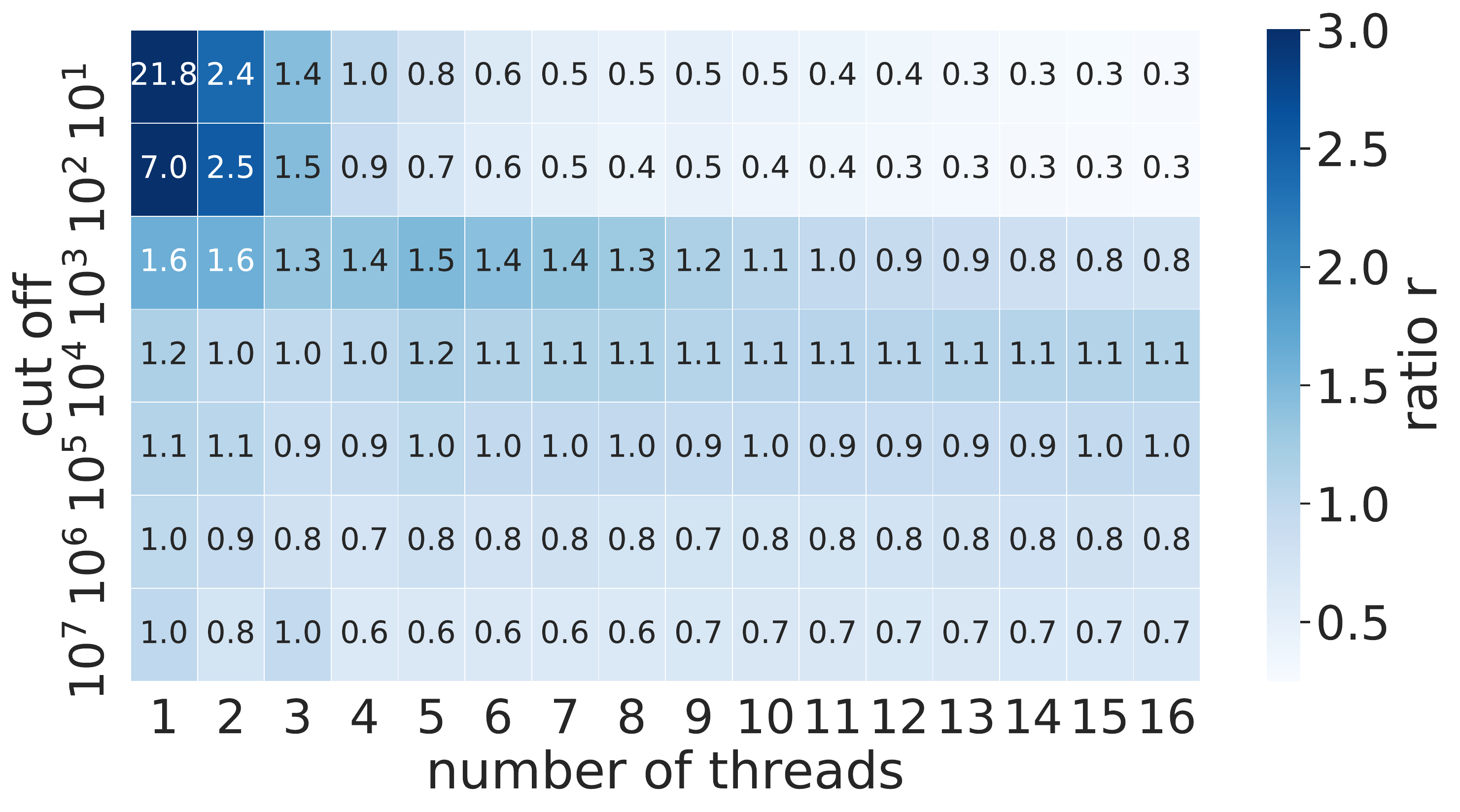}
		\caption{GOMP/hpxMP}
		\label{fig:ratio,sort,gcc}
	\end{subfigure}
	\caption{Speedup Ratio of Barcelona OpenMP Task suit's Sort Benchmark ran over several threads and cut off values using the hpxMP, llvm-OpenMP, and GOMP implementations. Values greater than $1$ mean that the OpenMP implementation achieved better performance when compared to hpxMP. Values below $1$ indicate that hpxMP outperformed the OpenMP implementation.}
	\label{fig:ratio,sort}
\end{figure}

\subsection{Blazemark}
In this section, the dmatdmatadd benchmark from Blaze's benchmark
suite{\footnote{\url{https://bitbucket.org/blaze-lib/blaze/wiki/Benchmarks}}}
is rerun to demonstrate the recent improvements in performance
when compared to the authors previous work~\cite{2019arXiv190303023Z} .
Blaze~\cite{blazelib} is a high performance C++ linear algebra
library which can use different backends for parallelization.
It also provides a benchmark suite called Blazemark for comparing
the performance of several linear algebra libraries, as well as different
backends used by Blaze, for a selection of arithmetic operations.
The results obtained from dmatdmatadd are presented and $4$
graphs are illustrated for a specific number of cores
($1$, $4$, $8$, and $16$) accordingly. The series in the graphs are
obtained by running the benchmark with llvm-OpenMP, an older version of
hpxMP, and the current state of hpxMP~\cite{2019arXiv190303023Z}.

\begin{figure}
	\centering
	\begin{subfigure}[b]{0.35\textwidth}
		\includegraphics[width=\linewidth]{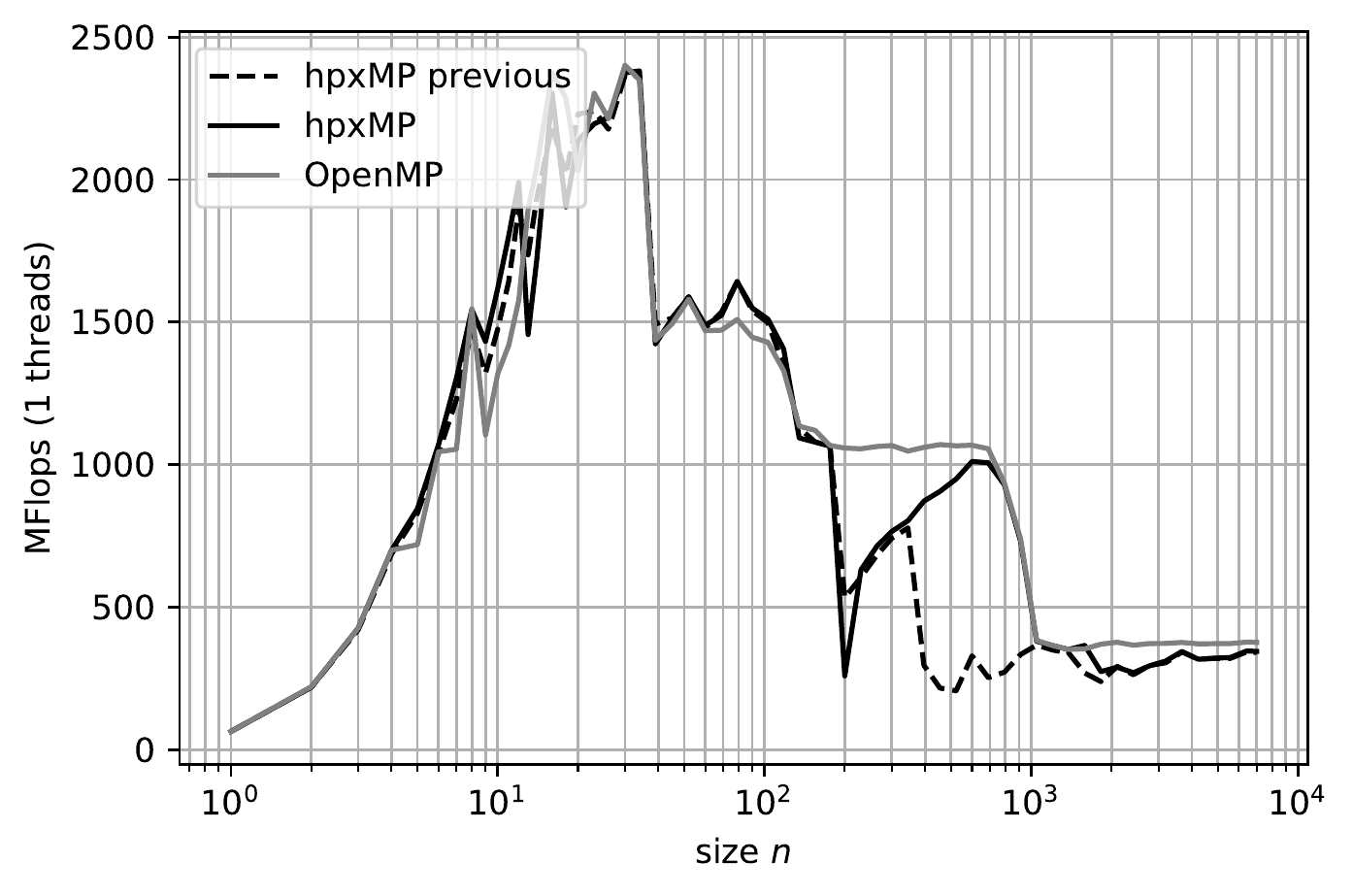}
		\caption{}
		\label{fig:ratio,dmatdmatadd_1}
	\end{subfigure}
	\begin{subfigure}[b]{0.35\textwidth}
		\includegraphics[width=\linewidth]{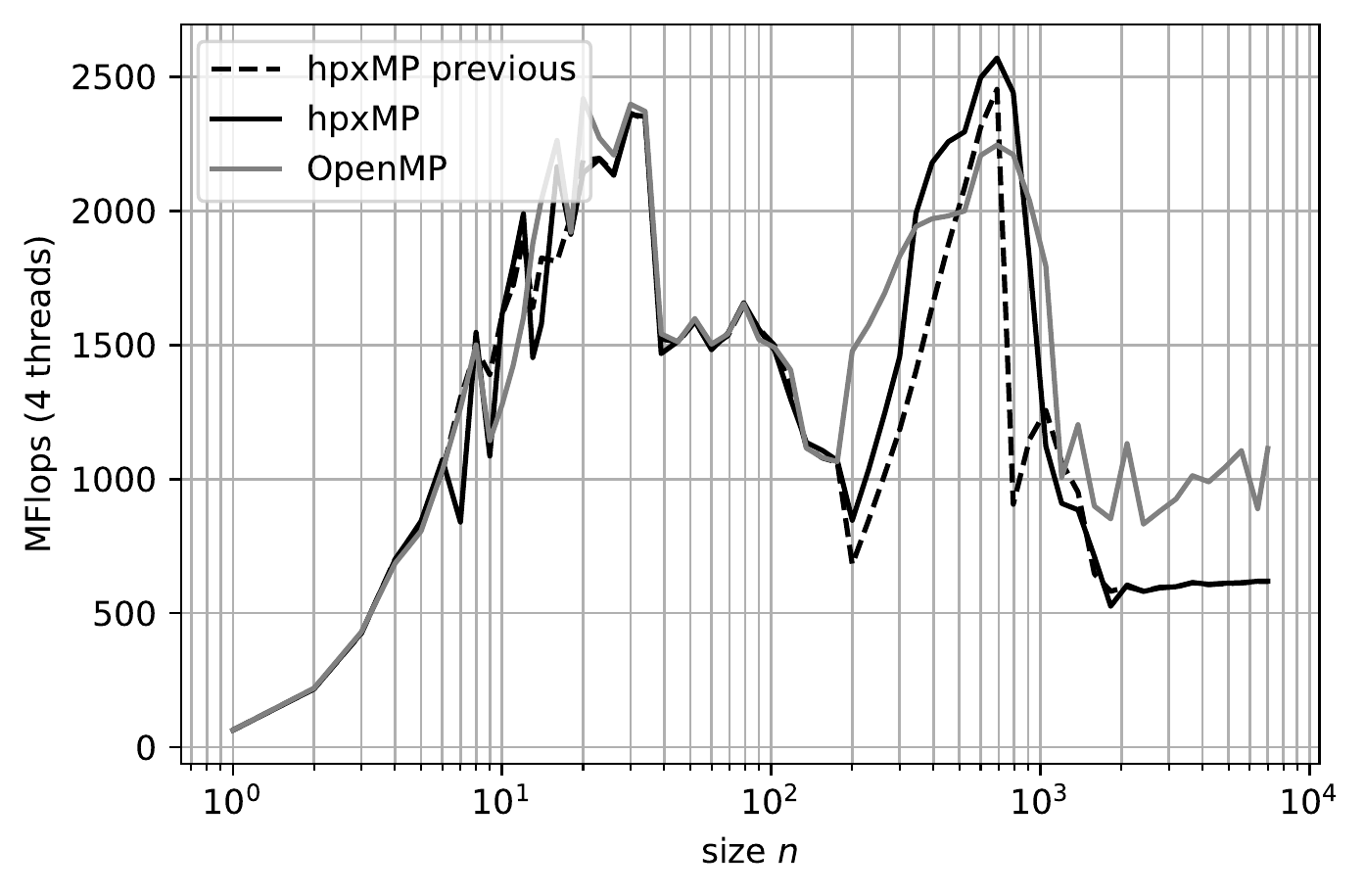}
		\caption{}
		\label{fig:ratio,dmatdmatadd_4}
	\end{subfigure}
	\begin{subfigure}[b]{0.35\textwidth}
		\includegraphics[width=\linewidth]{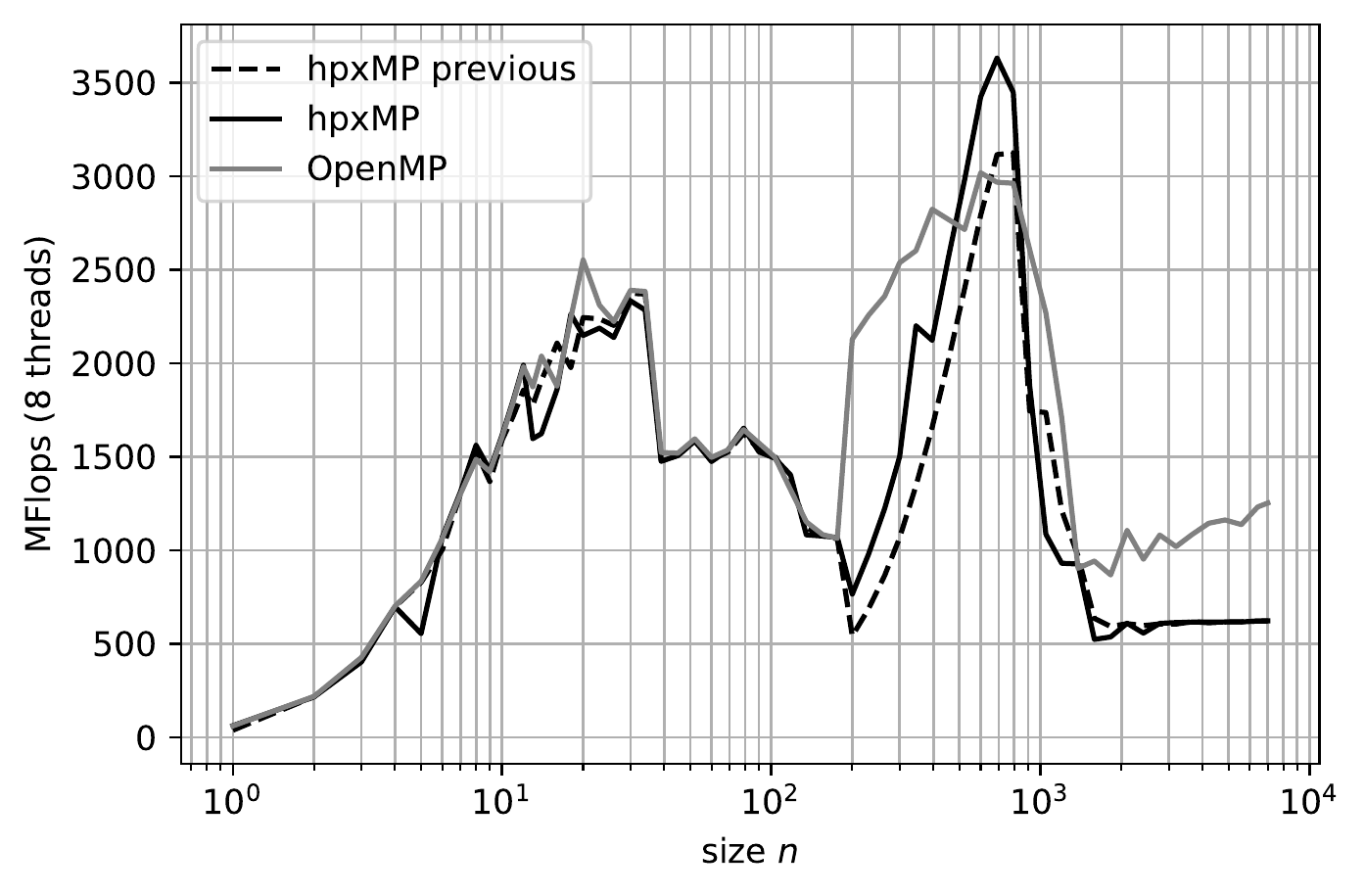}
		\caption{}
		\label{fig:ratio,dmatdmatadd_8}
	\end{subfigure}
	\begin{subfigure}[b]{0.35\textwidth}
		\includegraphics[width=\linewidth]{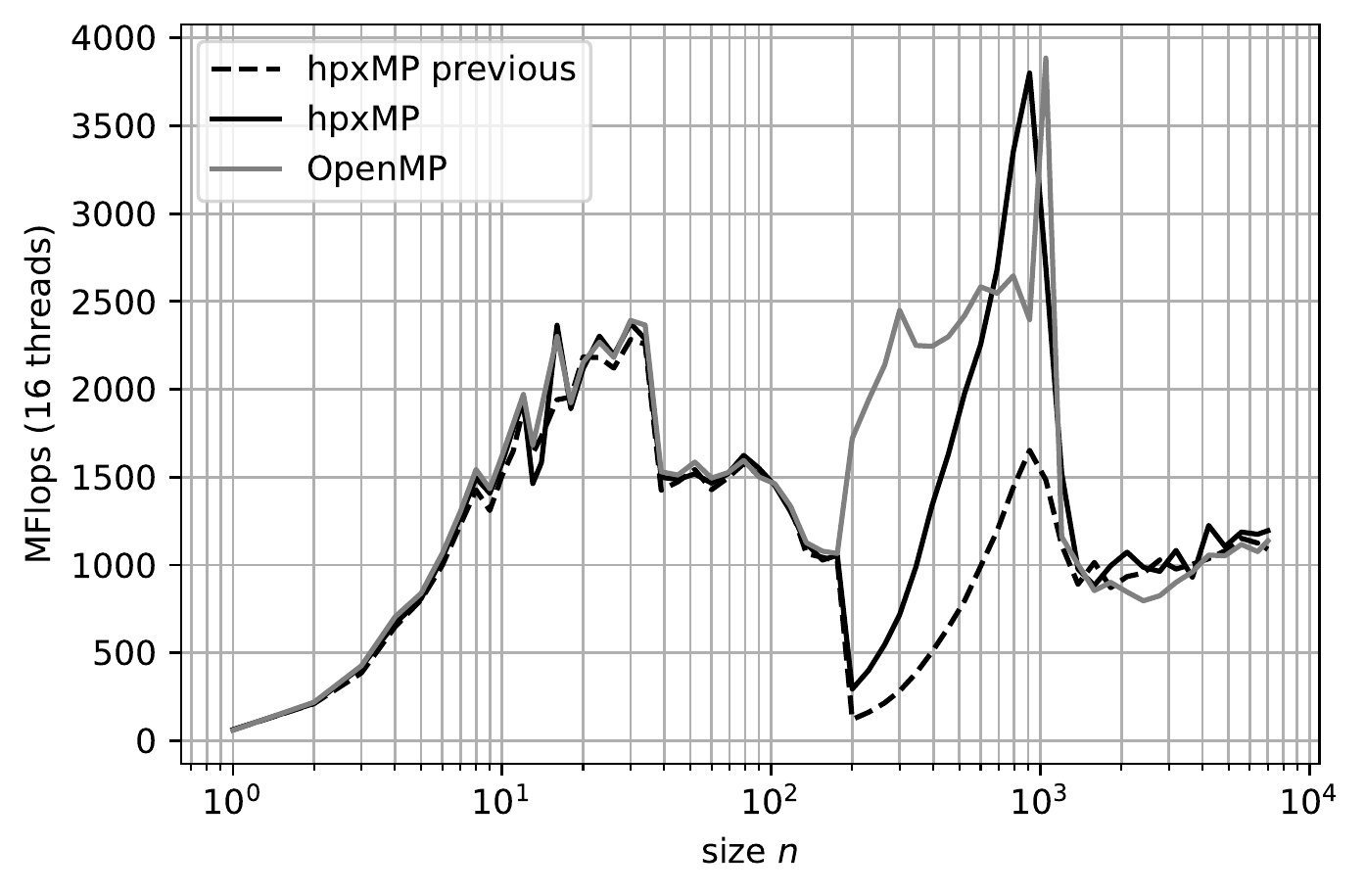}
		\caption{}
		\label{fig:ratio,dmatdmatadd_16}
	\end{subfigure}
	\caption{Scaling plots for dmatdmatadd Benchmarks for different number of threads, compared to llvm-OpenMP and previous work of the authors~\cite{2019arXiv190303023Z}: (a) $1$, (b) $4$, (c) $8$, and (d) $16$}
	\label{fig:scale,dmatdmatadd}
\end{figure}


The Dense Matrix Addition benchmark (dmatdmatadd) adds two dense
matrices $A$ and $B$, where $A,B\in \mathbb{R}^{n\times n}$, and writes the
result in matrix $C\in \mathbb{R}^{n\times n}$.
The operation can be written as $C[i,j]=A[i,j]+B[i,j]$.
Blaze uses the threshold of $36,100$ elements, which corresponds to
matrix size $190$ by $190$, before parallelizing the operation.
Matrices with less than $36,100$ elements are added sequentially.
Figure~\ref{fig:scale,dmatdmatadd} demonstrates the new scaling results
for the dmatdmatadd benchmark using $1$, $4$, $8$, and $16$ cores.
We observe notable improvement in performance between the previous
version of hpxMP and the current version. The performance
more closely mimic that of llvm-OpenMP.

\subsection{Discussion}
With the results presented above, we showed that hpxMP has similar performance to llvm-OpenMP for larger input sizes.
For some specific cases hpxMP was faster than llvm-OpenMP. This occurred because the operation of joining HPX
threads at the end of a parallel region introduces less overheads than the corresponding operation in llvm-OpenMP.
Joining the HPX threads are now done with a latch which is executed in user space. The cost of the operation amounts to a single atomic decrement per spawned
HPX thread. However, llvm-OpenMP uses kernel threads and therefore
must wait for the operating system to join the participating threads.

For smaller input sizes however, the hpxMP is less performant
as the overheads introduced by the HPX scheduler
are more significant compared to the actual workload. HPX threads
require their own stack segment as HPX threads are allowed to be suspended.
OpenMP does not incur this overhead as launched tasks are not able to be suspended.
In this way, the llvm-OpenMP implementation produces fewer scheduling overheads.


\section{Conclusion and Outlook}
\label{sec:outlook}
This work extended hpxMP, an implementation of the OpenMP standard utilizing the light-weight user level threads of HPX, with a subset og the task features of the OpenMP 5.0 standard. This contribution is one step towards the compatibility between OpenMP and AMT systems as it demonstrates a technique that enables AMT system to leverage highly-optimized OpenMP libraries. For the Barcelona OpenMP Task benchmark, hpxMP exhibited similar performance when compared to other OpenMP runtimes for large task sizes. However, it was not able to compete with these runtimes when faced with small tasks sizes. This performance decrement arises from the more general purpose threads created in HPX. For the \lstinline|#pragma omp parallel for| pragma, hpxMP has similar performance for larger input sizes. By using the HPX latch, the performance could be improved. These results show that hpxMP provides a way for bridging the compatibility gap between OpenMP and AMTs with acceptable performance for larger input sizes or larger task sizes.

In the future, we plan to improve performance for smaller input sizes by adding non-suspending threads to HPX, which do not require a stack, and thus reduce the overhead of thread creation and management. Additionally we plan to test the performance of HPX applications which use legacy OpenMP libraries, \emph{e.g.} Intel MKL. However, more of the OpenMP specification needs to be implemented within hpxMP. These experiments will serve as further validation of the techniques introduced in this paper.

\section*{Acknowledgment}
The work on hpxMP is funded by the
National Science Foundation (award $1737785$)
and and the Department of Defense
(DTIC Contract FA8075-14-D-0002/0007). Any opinions,
findings, conclusions or recommendations expressed in this material are
those of the authors and do not necessarily reflect the views of the National
Science Foundation or the Department of Defense.

\appendix
\section{Source code}
The source code of hpxMP~\cite{kemp2019hpxmp} and HPX is available on github\footnote{\url{https://github.com/STEllAR-GROUP/hpxMP},\url{https://github.com/STEllAR-GROUP/hpx}} released under the BSL-$1$.$0$.
%
\bibliographystyle{IEEEtran}
\bibliography{bib,hpx}

\begin{thebibliography}{10}
\providecommand{\url}[1]{#1}
\csname url@samestyle\endcsname
\providecommand{\newblock}{\relax}
\providecommand{\bibinfo}[2]{#2}
\providecommand{\BIBentrySTDinterwordspacing}{\spaceskip=0pt\relax}
\providecommand{\BIBentryALTinterwordstretchfactor}{4}
\providecommand{\BIBentryALTinterwordspacing}{\spaceskip=\fontdimen2\font plus
\BIBentryALTinterwordstretchfactor\fontdimen3\font minus
  \fontdimen4\font\relax}
\providecommand{\BIBforeignlanguage}[2]{{%
\expandafter\ifx\csname l@#1\endcsname\relax
\typeout{** WARNING: IEEEtran.bst: No hyphenation pattern has been}%
\typeout{** loaded for the language `#1'. Using the pattern for}%
\typeout{** the default language instead.}%
\else
\language=\csname l@#1\endcsname
\fi
#2}}
\providecommand{\BIBdecl}{\relax}
\BIBdecl

\bibitem{2019arXiv190303023Z}
\BIBentryALTinterwordspacing
T.~Zhang, S.~Shirzad, P.~Diehl, R.~Tohid, W.~Wei, and H.~Kaiser, ``An
  introduction to hpxmp: A modern openmp implementation leveraging hpx, an
  asynchronous many-task system,'' in \emph{Proceedings of the International
  Workshop on OpenCL}, ser. IWOCL'19.\hskip 1em plus 0.5em minus 0.4em\relax
  New York, NY, USA: ACM, 2019, pp. 13:1--13:10. [Online]. Available:
  \url{http://doi.acm.org/10.1145/3318170.3318191}
\BIBentrySTDinterwordspacing

\bibitem{thoman2018taxonomy}
P.~Thoman, K.~Dichev, T.~Heller, R.~Iakymchuk, X.~Aguilar, K.~Hasanov,
  P.~Gschwandtner, P.~Lemarinier, S.~Markidis, H.~Jordan \emph{et~al.}, ``A
  taxonomy of task-based parallel programming technologies for high-performance
  computing,'' \emph{The Journal of Supercomputing}, vol.~74, no.~4, pp.
  1422--1434, 2018.

\bibitem{heller2017hpx}
T.~Heller, P.~Diehl, Z.~Byerly, J.~Biddiscombe, and H.~Kaiser, ``Hpx an open
  source c++ standard library for parallelism and concurrency,'' 2017.

\bibitem{openmp08}
\BIBentryALTinterwordspacing
{OpenMP Architecture Review Board}, ``{OpenMP} application program interface
  version 3.0,'' May 2008. [Online]. Available:
  \url{http://www.openmp.org/mp-documents/spec30.pdf}
\BIBentrySTDinterwordspacing

\bibitem{openmp13}
\BIBentryALTinterwordspacing
------, ``{OpenMP} application program interface version 4.0,'' July 2013.
  [Online]. Available:
  \url{https://www.openmp.org/wp-content/uploads/OpenMP4.0.0.pdf}
\BIBentrySTDinterwordspacing

\bibitem{openmp15}
\BIBentryALTinterwordspacing
------, ``{OpenMP} application program interface version 4.5,'' November 2015.
  [Online]. Available:
  \url{https://www.openmp.org/wp-content/uploads/openmp-4.5.pdf}
\BIBentrySTDinterwordspacing

\bibitem{openmp18}
\BIBentryALTinterwordspacing
------, ``{OpenMP} application program interface version 5.0,'' November 2018.
  [Online]. Available:
  \url{https://www.openmp.org/wp-content/uploads/OpenMP-API-Specification-5.0.pdf}
\BIBentrySTDinterwordspacing

\bibitem{pthreads}
\BIBentryALTinterwordspacing
R.~A. Alfieri, ``An efficient kernel-based implementation of {POSIX} threads,''
  in \emph{Proceedings of the USENIX Summer 1994 Technical Conference on USENIX
  Summer 1994 Technical Conference - Volume 1}, ser. USTC'94.\hskip 1em plus
  0.5em minus 0.4em\relax Berkeley, CA, USA: USENIX Association, 1994, pp.
  5--5. [Online]. Available: \url{http://portal.acm.org/
  citation.cfm?id=1267257.1267262}
\BIBentrySTDinterwordspacing

\bibitem{inteltbb}
\BIBentryALTinterwordspacing
Intel, ``{Intel Thread Building Blocks},'' 2019,
  http://www.threadingbuildingblocks.org. [Online]. Available:
  \url{http://www.threadingbuildingblocks.org/}
\BIBentrySTDinterwordspacing

\bibitem{microsoftppl}
\BIBentryALTinterwordspacing
Microsoft, ``{Microsoft Parallel Pattern Library},'' 2010,
  http://msdn.microsoft.com/en-us/library/dd492418.aspx. [Online]. Available:
  \url{http://msdn.microsoft.com/en-us/library/dd492418.aspx}
\BIBentrySTDinterwordspacing

\bibitem{CarterEdwards20143202}
H.~C. Edwards, C.~R. Trott, and D.~Sunderland, ``Kokkos: Enabling manycore
  performance portability through polymorphic memory access patterns,''
  \emph{Journal of Parallel and Distributed Computing}, vol.~74, no.~12, pp.
  3202 -- 3216, 2014, domain-Specific Languages and High-Level Frameworks for
  High-Performance Computing.

\bibitem{chamberlain07parallelprogrammability}
B.~L. Chamberlain, D.~Callahan, and H.~P. Zima, ``{Parallel Programmability and
  the Chapel Language},'' \emph{International Journal of High Performance
  Computing Applications (IJHPCA)}, vol.~21, no.~3, pp. 291--312, 2007,
  \url{https://dx.doi.org/10.1177/1094342007078442}.

\bibitem{cilk++}
\BIBentryALTinterwordspacing
C.~E. Leiserson, ``{The Cilk++ concurrency platform},'' in \emph{DAC '09:
  Proceedings of the 46th Annual Design Automation Conference}.\hskip 1em plus
  0.5em minus 0.4em\relax New York, NY, USA: ACM, 2009, pp. 522--527. [Online].
  Available: \url{http://dx.doi.org/10.1145/1629911.1630048}
\BIBentrySTDinterwordspacing

\bibitem{openmp11}
\BIBentryALTinterwordspacing
{OpenMP Architecture Review Board}, ``{OpenMP} application program interface
  version 3.1,'' July 2011. [Online]. Available:
  \url{https://www.openmp.org/wp-content/uploads/OpenMP3.1.pdf}
\BIBentrySTDinterwordspacing

\bibitem{10.1109/MC.2016.232}
D.~A. Bader, ``Evolving mpi+x exascale,'' \emph{Computer}, vol.~49, no.~8,
  p.~10, 2016.

\bibitem{barrett2015toward}
R.~F. Barrett, D.~T. Stark, C.~T. Vaughan, R.~E. Grant, S.~L. Olivier, and
  K.~T. Pedretti, ``Toward an evolutionary task parallel integrated mpi+ x
  programming model,'' in \emph{Proceedings of the Sixth International Workshop
  on Programming Models and Applications for Multicores and Manycores}.\hskip
  1em plus 0.5em minus 0.4em\relax ACM, 2015, pp. 30--39.

\bibitem{rabenseifner2009hybrid}
R.~Rabenseifner, G.~Hager, and G.~Jost, ``Hybrid mpi/openmp parallel
  programming on clusters of multi-core smp nodes,'' in \emph{2009 17th
  Euromicro international conference on parallel, distributed and network-based
  processing}.\hskip 1em plus 0.5em minus 0.4em\relax IEEE, 2009, pp. 427--436.

\bibitem{smith2001development}
L.~Smith and M.~Bull, ``Development of mixed mode mpi/openmp applications,''
  \emph{Scientific Programming}, vol.~9, no. 2-3, pp. 83--98, 2001.

\bibitem{bak2017integrating}
S.~Bak, H.~Menon, S.~White, M.~Diener, and L.~Kale, ``Integrating openmp into
  the charm++ programming model,'' in \emph{Proceedings of the Third
  International Workshop on Extreme Scale Programming Models and
  Middleware}.\hskip 1em plus 0.5em minus 0.4em\relax ACM, 2017, p.~4.

\bibitem{agullo2017bridging}
E.~Agullo, O.~Aumage, B.~Bramas, O.~Coulaud, and S.~Pitoiset, ``Bridging the
  gap between openmp and task-based runtime systems for the fast multipole
  method,'' \emph{IEEE Transactions on Parallel and Distributed Systems},
  vol.~28, no.~10, pp. 2794--2807, 2017.

\bibitem{augonnet2011starpu}
C.~Augonnet, S.~Thibault, R.~Namyst, and P.-A. Wacrenier, ``Starpu: a unified
  platform for task scheduling on heterogeneous multicore architectures,''
  \emph{Concurrency and Computation: Practice and Experience}, vol.~23, no.~2,
  pp. 187--198, 2011.

\bibitem{gautier2007kaapi}
T.~Gautier, X.~Besseron, and L.~Pigeon, ``Kaapi: A thread scheduling runtime
  system for data flow computations on cluster of multi-processors,'' in
  \emph{Proceedings of the 2007 international workshop on Parallel symbolic
  computation}, 2007, pp. 15--23.

\bibitem{broquedis2012libkomp}
F.~Broquedis, T.~Gautier, and V.~Danjean, ``Libkomp, an efficient openmp
  runtime system for both fork-join and data flow paradigms,'' in
  \emph{International Workshop on OpenMP}.\hskip 1em plus 0.5em minus
  0.4em\relax Springer, 2012, pp. 102--115.

\bibitem{gautier2013xkaapi}
T.~Gautier, J.~V. Lima, N.~Maillard, and B.~Raffin, ``Xkaapi: A runtime system
  for data-flow task programming on heterogeneous architectures,'' in
  \emph{2013 IEEE 27th International Symposium on Parallel and Distributed
  Processing}.\hskip 1em plus 0.5em minus 0.4em\relax IEEE, 2013, pp.
  1299--1308.

\bibitem{hpx_pgas_2014}
H.~Kaiser, T.~Heller, B.~A. Lelbach, A.~Serio, and D.~Fey, ``{HPX: A Task Based
  Programming Model in a Global Address Space},'' in \emph{Proceedings of the
  International Conference on Partitioned Global Address Space Programming
  Models (PGAS)}, ser. art. id 6, 2014,
  \url{https://stellar.cct.lsu.edu/pubs/pgas14.pdf}.

\bibitem{heller:2012}
T.~Heller, H.~Kaiser, and K.~Iglberger, ``{Application of the ParalleX
  Execution Model to Stencil-Based Problems},'' \emph{Computer Science -
  Research and Development}, vol.~28, no. 2-3, pp. 253--261, 2012,
  \url{https://stellar.cct.lsu.edu/pubs/isc2012.pdf}.

\bibitem{Heller:2013:UHL:2530268.2530269}
T.~Heller, H.~Kaiser, A.~Sch\"{a}fer, and D.~Fey, ``{Using HPX and LibGeoDecomp
  for Scaling HPC Applications on Heterogeneous Supercomputers},'' in
  \emph{Proceedings of the ACM/IEEE Workshop on Latest Advances in Scalable
  Algorithms for Large-Scale Systems (ScalA, SC Workshop)}, ser. art. id 1,
  2013, \url{https://stellar.cct.lsu.edu/pubs/scala13.pdf}.

\bibitem{Kaiser:2015:HPL:2832241.2832244}
H.~Kaiser, T.~Heller, D.~Bourgeois, and D.~Fey, ``{Higher-level Parallelization
  for Local and Distributed Asynchronous Task-based Programming},'' in
  \emph{Proceedings of the ACM/IEEE International Workshop on Extreme Scale
  Programming Models and Middleware (ESPM, SC Workshop)}, 2015, pp. 29--37,
  \url{https://stellar.cct.lsu.edu/pubs/executors_espm2_2015.pdf}.

\bibitem{hartmut_kaiser_2018_1484427}
\BIBentryALTinterwordspacing
H.~Kaiser, B.~A.~L. aka wash, T.~Heller, A.~Bergé, J.~Biddiscombe, and M.~S.
  et.al., ``{STEllAR-GROUP/hpx: HPX V1.2.0: The C++ Standards Library for
  Parallelism and Concurrency},'' Nov. 2018. [Online]. Available:
  \url{https://doi.org/10.5281/zenodo.1484427}
\BIBentrySTDinterwordspacing

\bibitem{heller:hpc_2016}
T.~Heller, H.~Kaiser, P.~Diehl, D.~Fey, and M.~A. Schweitzer, ``Closing the
  performance gap with modern c++,'' in \emph{High Performance Computing}, ser.
  Lecture Notes in Computer Science, M.~Taufer, B.~Mohr, and J.~M. Kunkel,
  Eds., vol. 9945.\hskip 1em plus 0.5em minus 0.4em\relax Springer
  International Publishing, 2016, pp. 18--31.

\bibitem{wagle2018methodology}
B.~Wagle, S.~Kellar, A.~Serio, and H.~Kaiser, ``Methodology for adaptive active
  message coalescing in task based runtime systems,'' in \emph{2018 IEEE
  International Parallel and Distributed Processing Symposium Workshops
  (IPDPSW)}.\hskip 1em plus 0.5em minus 0.4em\relax IEEE, 2018, pp. 1133--1140.

\bibitem{scaling_impaired_apps}
H.~Kaiser, M.~Brodowicz, and T.~Sterling, ``{ParalleX}: An advanced parallel
  execution model for scaling-impaired applications,'' in \emph{Parallel
  Processing Workshops}.\hskip 1em plus 0.5em minus 0.4em\relax Los Alamitos,
  CA, USA: IEEE Computer Society, 2009, pp. 394--401.

\bibitem{wagle2019runtime}
\BIBentryALTinterwordspacing
B.~Wagle, M.~A.~H. Monil, K.~Huck, A.~D. Malony, A.~Serio, and H.~Kaiser,
  ``Runtime adaptive task inlining on asynchronous multitasking runtime
  systems,'' in \emph{Proceedings of the 48th International Conference on
  Parallel Processing}, ser. ICPP 2019.\hskip 1em plus 0.5em minus 0.4em\relax
  New York, NY, USA: ACM, 2019, pp. 76:1--76:10. [Online]. Available:
  \url{http://doi.acm.org/10.1145/3337821.3337915}
\BIBentrySTDinterwordspacing

\bibitem{7040972}
R.~F. {Van der Wijngaart} and T.~G. {Mattson}, ``The parallel research
  kernels,'' in \emph{2014 IEEE High Performance Extreme Computing Conference
  (HPEC)}, Sep. 2014, pp. 1--6.

\bibitem{akl1987optimal}
S.~G. Akl and N.~Santoro, ``Optimal parallel merging and sorting without memory
  conflicts,'' \emph{IEEE Transactions on Computers}, vol. 100, no.~11, pp.
  1367--1369, 1987.

\bibitem{gautier2018impact}
T.~Gautier, C.~P{\'e}rez, and J.~Richard, ``On the impact of openmp task
  granularity,'' in \emph{International Workshop on OpenMP}.\hskip 1em plus
  0.5em minus 0.4em\relax Springer, 2018, pp. 205--221.

\bibitem{vandierendonck2011unified}
H.~Vandierendonck, G.~Tzenakis, and D.~S. Nikolopoulos, ``A unified scheduler
  for recursive and task dataflow parallelism,'' in \emph{2011 International
  Conference on Parallel Architectures and Compilation Techniques}.\hskip 1em
  plus 0.5em minus 0.4em\relax IEEE, 2011, pp. 1--11.

\bibitem{bueno2011productive}
J.~Bueno, L.~Martinell, A.~Duran, M.~Farreras, X.~Martorell, R.~M. Badia,
  E.~Ayguade, and J.~Labarta, ``Productive cluster programming with ompss,'' in
  \emph{European Conference on Parallel Processing}.\hskip 1em plus 0.5em minus
  0.4em\relax Springer, 2011, pp. 555--566.

\bibitem{blazelib}
K.~{Iglberger}, ``Blaze c++ linear algebra library,''
  https://bitbucket.org/blaze-lib, 2012.

\bibitem{kemp2019hpxmp}
\BIBentryALTinterwordspacing
J.~Kemp, T.~Zhang, S.~Shirzad, B.~A.~L. aka wash, H.~Kaiser, B.~Wagle,
  P.~Amini, and A.~Kheirkhahan, ``{STEllAR-GROUP/hpxMP: hpxMP v0.3.0: An OpenMP
  runtime implemented using HPX},'' Nov. 2019. [Online]. Available:
  \url{https://doi.org/10.5281/zenodo.3554559}
\BIBentrySTDinterwordspacing

\end{thebibliography}

\end{document}